\documentclass[preprint,12pt]{elsarticle}
\journal{Journal of Sound and Vibration}
\usepackage{graphics,epsfig}
\usepackage{graphicx}
\usepackage{float}
\usepackage{amssymb,amstext,amsmath}
\usepackage{mathtools}
\mathtoolsset{showonlyrefs}
\usepackage{booktabs}
\usepackage{natbib}
\DeclareMathOperator*{\argmax}{arg\,max}
\begin{document}
\begin{frontmatter}
\title{Optimal parameters uncoupling vibration modes of oscillators}
\author{K. C. Le\footnote{Corresponding author: phone: +49 234 32-26033, email: chau.le@rub.de.}, A. Pieper}
\address{Lehrstuhl f\"{u}r Mechanik - Materialtheorie, Ruhr-Universit\"{a}t Bochum,\\D-44780 Bochum, Germany}
\begin{abstract} 
A novel optimization concept for an oscillator with two degrees of freedom is proposed. By using specially defined motion ratios, we control the action of springs and dampers to each degree of freedom of the oscillator. If the potential action of the springs in one period of vibration, used as the payoff function for the conservative oscillator, is maximized, then the optimal motion ratios uncouple vibration modes. The same result holds true for the dissipative oscillator. The application to optimal design of vehicle suspension is discussed.
\end{abstract}


\end{frontmatter}

\section{Introduction}

In engineering praxis a vibration isolator is often required to reduce the transmission of forces or displacements to special bodies, mountings, or bearings while the system is excited. If the vibration of the bodies remains small and well controlled around a desired position of equilibrium for most of excitations, a comfortable, light, and durable system is created. The optimal design of vibration isolator can then be realized depending on the specific goal expressed in terms of the so-called payoff (or objective) function \citep{haug1979applied,arora2004introduction}. The fact that spring forces depend on displacements, and damping forces on velocities, often entice engineers to design a vibration isolator whose elements, springs and/or dampers, are positioned at the places of putative large relative displacements (or velocities) of the bodies. However, it turns out that for the oscillators having several degrees of freedom and modes of vibration, this does not always leads to the optimal solution.

What is said above can at best be illustrated on the practical example of a conventional cars suspension. Because large relative motions between the wheels and the chassis are visible, it seams that a position next to each wheel is the best for springs and dampers to be placed \citep{rowell1922principles,guest1926main,milliken1995race}. Due to the complexity of the optimization problem many authors used a quarter car model for the optimization purpose (see \citep{alkhatib2004optimal,turkay2005study,sun2007genetic} and the references therein). Since in this case the motion of the system is one-dimensional, all springs and dampers act in the direction of motion and their configuration is fixed. Thus, only the spring rates and damper constants can be varied in this optimization. With the goal of maximizing isolation of the chassis from a harmonic base excitation in the frequency domain to achieve the best ride quality of the vehicle, \citet{alkhatib2004optimal} used the root mean square of acceleration or displacement of the chassis as the payoff function. If the interest is in contrary to minimize the dynamic tire load, then the variance of the dynamic load used by \citet{sun2007genetic} serves as the payoff function. The optimization using a half-car model considered for instance by  \citet{tamboli1999optimum,giua2000mixed,sun2002optimum} and a full-car model by \citet{jayachandran2013modeling} deals again with fixed configurations of springs and dampers while varying their characteristics to meet similar goals. Note, however, that the fixing of special configuration of springs and dampers often exhibits some deficiency in damping of roll vibrations of conventional vehicle suspensions as shown by \citet{le2014damping} in an analysis of forced vibration using a half-car model. The first step in modifying this design concept of suspension by introducing a smart mechanism that adapts the installation ratios of both springs and dampers to different modes of vibrations in equal way has been proposed by \citet{pieper2014optimisation}. Nowadays, especially in tuning of vehicle suspension elements, a huge effort is spend on lap time simulations using different numerical packages \citep{kelly2010time}. An advanced approach is to measure the real-time motions on a specified system and control it by active springs and dampers. In this case the physical property of each element can be changed immediately and the optimal control is done by software and actuators at each time instant \citep{hac1985suspension,karnopp1989analytical,karnopp1995active,yamashita1994application,gao2006multi,sun2011finite}. However, this approach only allows an optimization after bad motions have already been detected. The common feature of traditional optimization of passive or active suspensions is that the concept of the dynamic system including the configuration of springs and dampers is fixed at the beginning and only the physical properties of the elements are subject to variation. Independent from the choice of payoff function, this optimization practice limits strongly the variability of dynamical system for comparison to  select the overall best solution. 

This paper focuses on a new optimization concept for an oscillator with the configuration of springs and dampers being subject to variation. This is realized by a mechanism (rocker) having several motion ratios controlling the action of springs and dampers to each degree of freedom. The variation of motion ratios allows to change the maximum force, induced by springs or
dampers, to different modes of vibration. Note that this optimization concept is close to that of topology optimization of materials \citep{bendsoe2013topology,junker2015variational,junker2016discontinuous} or optimization of placement of piezo-patches in smart structures \citep{crawley1987use,dhuri2009multi}. The springs get used most effectively if the spring energies (and consequently the magnitude of spring forces) are maximal when acting against the corresponding modes of vibration. This leads to the maximum of the total potential action of all springs over one period of each vibration mode. The same can be said in the case of dissipative oscillators with springs and dampers. The aim of this paper is to show that, if the potential action of the springs over one (conditional) period of vibration is used as the payoff function to be maximized, then the optimal parameters controlling the action of springs uncouple modes of vibrations and the maximum available forces of springs act against the normal modes. 

In order to prove this statement rigorously we need to apply the theory of optimal control processes \citep{pontryagin1987mathematical,bryson1975applied,athans2013optimal} to the special case of time-independent control parameters. In this case we are dealing with the variational problem with constraint imposed on the state variables of the dynamical system in form of the equation of motion depending on the time-independent control parameters. We formulate the extended Pontryagin's maximum principle and, alternatively, the necessary and sufficient conditions for the optimal control parameters of oscillators obeying the equations of small amplitude vibrations. We then apply this theory to the oscillator having two degrees of freedom, first with springs, and then later with springs and dampers, to prove the above statements. 

The paper is organized as follows. In the next Section we present the theory of optimal control parameters for oscillators. Sections 3 and 4 apply this theory to the conservative and dissipative oscillators, respectively. Finally, Section 5 discusses the optimal design concept and concludes the paper.

\section{Theory of optimal control parameters}

We let $m$-dimensional vector $\mathbf{a}=(a_1,\ldots ,a_m)$ denote time-independent control parameters of a mechanical system under consideration and assume that $\mathbf{a} \in \mathcal{A}\subseteq \mathbb{R}^m$, with $\mathcal{A}$ being the set of admissible control parameters. The motion of this mechanical system is governed by the equations
\begin{equation}
\label{2.1}
\begin{cases}
   \dot{\mathbf{x}}(t) = \mathbf{f}(\mathbf{x}(t),\mathbf{a}) \quad (t\ge 0), \\
   \mathbf{x}(0)=\mathbf{x}^0,
\end{cases}
\end{equation}
with $\mathbf{f}:\mathbb{R}^n\times \mathcal{A}\to \mathbb{R}^n$. We introduce the payoff function
\begin{equation}
\label{2.2}
P(\mathbf{a})=\int_0^T r(\mathbf{x}(t),\mathbf{a})\, dt+g(\mathbf{x}(T)),
\end{equation}
where the end-time $T>0$, running payoff $r:\mathbb{R}^n\times \mathcal{A}\to \mathbb{R}$ and end-time payoff $g:\mathbb{R}^n\to \mathbb{R}$ are given. The problem is to find optimal parameters $\mathbf{a}^*$ that maximize payoff function \eqref{2.2} among all admissible $\mathbf{a}\in \mathcal{A}$ and $\mathbf{x}(t)$ satisfying constraints \eqref{2.1}. Note that the control parameters $\mathbf{a}=(a_1,\ldots ,a_m)$ can be identified with the control processes $\mathbf{u}(t)=(u_1(t),\ldots ,u_m(t))$ satisfying the constraints
\begin{equation}
\label{2.2a}
\begin{cases}
   \dot{\mathbf{u}}(t) = \mathbf{0} \quad (t\ge 0), \\
   \mathbf{u}(0)=\mathbf{a}.
\end{cases}
\end{equation}
Thus, the above formulated problem is the special case of the problem considered in the theory of optimal control processes \citep{pontryagin1987mathematical,bryson1975applied,athans2013optimal}.

Using the similarity with the problem of finding optimal control processes solved by \citet{boltyansky1956theory}, we introduce the control theory Hamiltonian
\begin{equation}
\label{2.3}
H(\mathbf{x},\mathbf{p},\mathbf{a})=\mathbf{p}\cdot \mathbf{f}(\mathbf{x},\mathbf{a})+r(\mathbf{x},\mathbf{a}), \quad (\mathbf{x},\mathbf{p}\in \mathbb{R}^n, \mathbf{a}\in \mathcal{A}).
\end{equation}
Pontryagin's maximum principle can be extended to the problem of finding optimal control parameters in form of the following theorem: Assume $\mathbf{a}^*$ is optimal for \eqref{2.1}, \eqref{2.2} and $\mathbf{x}^*(t)$ is the corresponding motion. Then there exists a function $\mathbf{p}^*(t)$ such that
\begin{equation}
\label{2.4}
\begin{split}
\dot{\mathbf{x}}^*(t)=H_\mathbf{p}(\mathbf{x}^*(t),\mathbf{p}^*(t),\mathbf{a}^*),
\\
\dot{\mathbf{p}}^*(t)=-H_\mathbf{x}(\mathbf{x}^*(t),\mathbf{p}^*(t),\mathbf{a}^*),
\end{split}
\end{equation}
where $H_\mathbf{p}$ and $H_\mathbf{x}$ are the partial derivatives of $H$ with respect to $\mathbf{p}$ and $\mathbf{x}$, respectively, and
\begin{equation}
\label{2.5}
\mathbf{a}^*=\argmax_{\mathbf{a}\in \mathcal{A}}\int_0^TH(\mathbf{x}^*(t),\mathbf{p}^*(t),\mathbf{a})\, dt.
\end{equation}
In addition, $H(\mathbf{x}^*(t),\mathbf{p}^*(t),\mathbf{a}^*)$ remains constant and $\mathbf{p}^*(t)$ satisfies the end condition
\begin{equation}
\label{2.6}
\mathbf{p}^*(T)=g_\mathbf{x}(\mathbf{x}^*(T)).
\end{equation}
In contrast to the maximum principle for the optimal control processes (cf. \citep{pontryagin1987mathematical}), condition \eqref{2.5} indicates that the optimal control parameters must be found from maximizing the {\it integral} of Hamiltonian among all admissible control parameters. We omit the proof of this theorem which is quite similar to that given in the theory of optimal control processes \citep{pontryagin1987mathematical}.

For the more specific case of oscillators depending on the control parameters $\mathbf{a}=(a_1,\ldots ,a_m)\in \mathcal{A}$ it is convenient to use the alternative form of the maximum principle. The small amplitude free vibration of an oscillator having $n$ degrees of freedom is governed by the system (see, e.g., \citep{le2014energy})
\begin{equation}
\label{2.9}
\begin{cases}
\mathbf{M}(\mathbf{a})\ddot{\mathbf{q}}(t)+\mathbf{C}(\mathbf{a})\dot{\mathbf{q}}(t)+\mathbf{K}(\mathbf{a})\mathbf{q}(t)=\mathbf{0},\quad (t>0), \\
   \mathbf{q}(0)=\mathbf{0},\quad \dot{\mathbf{q}}(0)=\mathbf{v}^0
\end{cases}
\end{equation}
Here $\mathbf{q}(t)=(q_1(t),\ldots ,q_n(t))^T$, while $\mathbf{M}(\mathbf{a})$, $\mathbf{C}(\mathbf{a})$, and $\mathbf{K}(\mathbf{a})$ are mass, damping, and stiffness symmetric $n\times n$-matrix, respectively. Our aim is to maximize the payoff function
\begin{equation}
\label{2.10}
P(\mathbf{a})=\int_0^T r(\mathbf{q}(t),\dot{\mathbf{q}}(t),\mathbf{a})\, dt
\end{equation}
among all admissible parameters $\mathbf{a}\in \mathcal{A}$ and $\mathbf{q}(t)$ satisfying constraint \eqref{2.9}. Following the same strategy, we introduce the control theory Lagrangian
\begin{equation}
\label{2.11}
L(\mathbf{q},\dot{\mathbf{q}},\ddot{\mathbf{q}},\mathbf{p},\mathbf{a})=\mathbf{p}\cdot [\mathbf{M}(\mathbf{a})\ddot{\mathbf{q}}+\mathbf{C}(\mathbf{a})\dot{\mathbf{q}}+\mathbf{K}(\mathbf{a})\mathbf{q}]+r(\mathbf{q},\dot{\mathbf{q}},\mathbf{a}), 
\end{equation}
where $\mathbf{q},\dot{\mathbf{q}},\ddot{\mathbf{q}},\mathbf{p}\in \mathbb{R}^n$, $\mathbf{a}\in \mathcal{A}$. The dual quantity $\mathbf{p}$ plays the role of the Lagrange multiplier that enables one to get rid of constraint \eqref{2.9}.

The maximum principle for optimal control parameters can be formulated in this case as follows. Assume $\mathbf{a}^*$ is optimal for \eqref{2.9}, \eqref{2.10} and $\mathbf{q}^*(t)$ is the corresponding motion, with $\dot{\mathbf{q}}^*(t)$ and $\ddot{\mathbf{q}}^*(t)$ being the velocity and acceleration, respectively. Then there exists a vector-valued function $\mathbf{p}^*(t)$ such that
\begin{align}
\mathbf{M}(\mathbf{a}^*)\ddot{\mathbf{q}}^*(t)+\mathbf{C}(\mathbf{a}^*)\dot{\mathbf{q}}^*(t)+\mathbf{K}(\mathbf{a}^*)\mathbf{q}^*(t)&=\mathbf{0}, \label{2.12a}
\\
\mathbf{M}(\mathbf{a}^*)\ddot{\mathbf{p}}^*(t)-\mathbf{C}(\mathbf{a}^*)\dot{\mathbf{p}}^*(t)+\mathbf{K}(\mathbf{a}^*)\mathbf{p}^*(t)&=\left. \left( \frac{d}{dt}\frac{\partial r}{\partial \dot{\mathbf{q}}}-\frac{\partial r}{\partial \mathbf{q}}\right) \right|_{\mathbf{q}(t)=\mathbf{q}^*(t)},
\label{2.12b}
\end{align}
where the right-hand side of \eqref{2.12b} can be regarded as the external excitation for $\mathbf{p}^*(t)$. These equation are subjected to the initial and end conditions
\begin{equation}
\label{2.14}
\mathbf{q}^*(0)=0,\quad \dot{\mathbf{q}}^*(0)=\mathbf{v}^0,\quad \mathbf{p}^*(T)=0,\quad \dot{\mathbf{p}}^*(T)=0.
\end{equation}
Finally, the optimal parameters must be found from the following maximization problem
\begin{equation}
\label{2.13}
\mathbf{a}^*=\argmax_{\mathbf{a}\in \mathcal{A}}\int_0^T L(\mathbf{q}^*(t),\dot{\mathbf{q}}^*(t),\ddot{\mathbf{q}}^*(t),\mathbf{p}^*(t),\mathbf{a})\, dt.
\end{equation}
Note that, if some matrix in $L$ does not depend on $\mathbf{a}$, the corresponding term can be dropped in this maximization problem. Sometimes it is more convenient to verify the optimality of the expected control parameters by using the necessary conditions
\begin{equation}
\label{2.15}
\int_0^T L_{\mathbf{a}}(\mathbf{q}^*(t),\dot{\mathbf{q}}^*(t),\ddot{\mathbf{q}}^*(t),\mathbf{p}^*(t),\mathbf{a}^*)\, dt=0
\end{equation}
together with the sufficient condition that the matrix of second derivatives
\begin{equation}
\label{2.15a}
\int_0^T L_{\mathbf{a}\mathbf{a}}(\mathbf{q}^*(t),\dot{\mathbf{q}}^*(t),\ddot{\mathbf{q}}^*(t),\mathbf{p}^*(t),\mathbf{a}^*)\, dt
\end{equation}
is negative definite. Note that these conditions guarantee only the local maximum of the payoff function.

If there are additional constraints imposed on the admissible parameters in the form
\begin{equation}
\label{2.16}
f_\alpha (\mathbf{a})=0, \quad \alpha =1,\ldots ,j,
\end{equation}
then the Lagrangian must be modified to
\begin{equation}
\label{2.17}
L(\mathbf{q},\dot{\mathbf{q}},\ddot{\mathbf{q}},\mathbf{p},\mathbf{a},\lambda _\alpha )=\mathbf{p}\cdot [\mathbf{M}(\mathbf{a})\ddot{\mathbf{q}}+\mathbf{C}(\mathbf{a})\dot{\mathbf{q}}+\mathbf{K}(\mathbf{a})\mathbf{q}]+r(\mathbf{q},\dot{\mathbf{q}},\mathbf{a})-\sum_{\alpha =1}^j \lambda _\alpha f_\alpha (\mathbf{a}),
\end{equation}
with $\lambda _\alpha $ being the Lagrange multipliers for the constraints \eqref{2.16}. Alternatively, the set of constraints \eqref{2.16} determines a $(m-j)$-dimensional surface in the space of admissible parameters, so, by introducing the coordinates on this surface, we reduce the problem to the unconstrained maximization.

\section{Application to conservative oscillators}
In this Section we illustrate the application of the theory proposed in the previous Section on two examples of conservative oscillators having one and two degrees of freedom.

\begin{figure}[htbp]
	\centering
		\includegraphics{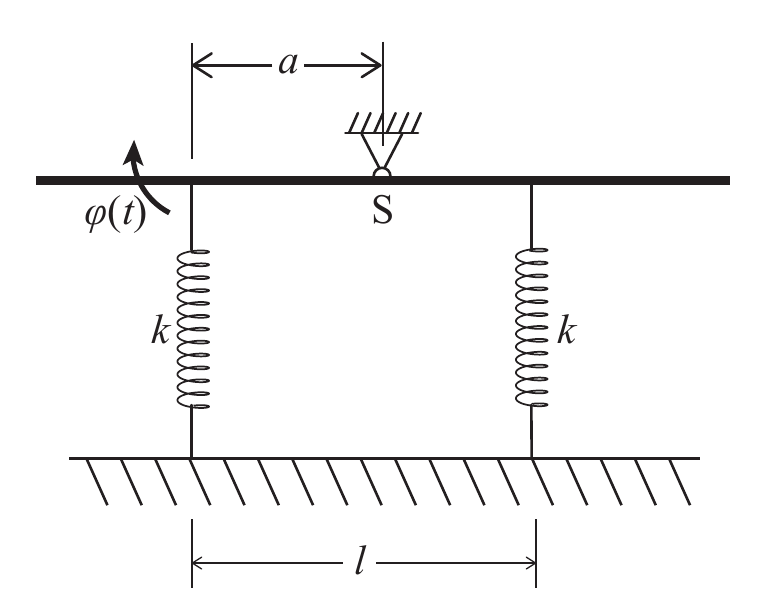}
	\caption{Rigid beam rotating about the fixed support O}
	\label{fig1}
\end{figure}

As the first simple example we consider a rigid beam rotating about the fixed support coinciding with its center of mass S and being connected with two springs of equal stiffness $k$ as shown in Fig.~\ref{fig1}. We fix the distance between the springs, $l$, as well as the spring stiffness $k$, but variate the placement of the springs. Thus, the free parameter controlling this oscillator is the distance $a$ between the first spring and S. We assume that $a\in (0,l)$. The only degree of freedom is the angle of rotation $\varphi $. The kinetic and potential energy of the oscillator read
\begin{equation}
\label{3.1}
K(\dot{\varphi})=\frac{1}{2}J_S\dot{\varphi }^2,\quad
U(\varphi)=\frac{1}{2}k[a^2\varphi ^2+(l-a)^2\varphi ^2],
\end{equation}
with $J_S$ denoting the moment of inertia (about S) of the beam. The free vibration of this oscillator is governed by
\begin{equation}
\label{3.2}
\begin{cases}
J_S\ddot{\varphi }+k[a^2+(l-a)^2]\varphi =0,\quad (t>0), \\
\varphi (0)=0,\quad \dot{\varphi }(0)=v^0.
\end{cases}
\end{equation}
The problem is to find the optimal parameter $a^*$ maximizing the potential action of the springs 
\begin{equation}
\label{3.3}
P(a)=\int_0^T \frac{1}{2}k[a^2\varphi ^2+(l-a)^2\varphi ^2]\, dt
\end{equation}
among all admissible $a\in (0,l)$ and $\varphi (t)$ satisfying constraint \eqref{3.2}.

In accordance with the theory of optimal control parameters proposed in the previous Section we construct the Lagrangian of the control theory
\begin{equation}
L(\varphi,\ddot{\varphi},p,a)=p[J_S\ddot{\varphi }+k(a^2+(l-a)^2)\varphi ] 
+\frac{1}{2}k[a^2+(l-a)^2]\varphi ^2. \label{3.4} 
\end{equation}
For the optimal control parameter $a^*$ we have to satisfy
\begin{gather}
J_S\ddot{\varphi }^*+k[a^{*2}+(l-a^*)^2]\varphi ^*=0, \notag
\\
J_S\ddot{p}^*+k[a^{*2}+(l-a^*)^2]p^*=-k[a^{*2}+(l-a^*)^2]\varphi ^*,
\label{3.6}
\\
\varphi ^*(0)=0,\quad \dot{\varphi }^*(0)=v^0,\quad p^*(T)=0, \quad \dot{p}^*(T)=0, \notag
\end{gather}
and
\begin{equation}
a^*=\argmax_{a\in (0,l)}k[a^2+(l-a)^2]\int_0^T (p^*\varphi ^*
+\frac{1}{2}\varphi ^{*2}) \, dt. \label{3.8}
\end{equation}
In \eqref{3.8} the first term of the Lagrangian \eqref{3.4} is omitted as independent of $a$. Since the expression in square brackets is positive definite and quadratic in $a$, the maximum of \eqref{3.8} is achieved at $a^*=l/2\in (0,l)$ provided
\begin{equation}
\label{3.10}
\int_0^T (p^*\varphi ^*+\frac{1}{2}\varphi ^{*2}) \, dt<0.
\end{equation}
To check condition \eqref{3.10} we must find $\varphi ^*(t)$ and $p^*(t)$ from \eqref{3.6} with  $a^*=l/2$. The solution for $\varphi ^*(t)$ reads
\begin{equation}
\label{3.10a}
\varphi ^*(t)=\frac{v^0}{\omega } \sin \omega t,
\end{equation}
where $\omega =\sqrt{kl^2/2J_S}$. Let us choose $T$ to be the period of vibration for this parameter $a^*=l/2$, so $T=2\pi /\omega $. Solving \eqref{3.6} with $p^*(T)=\dot{p}^*(T)=0$, we find
\begin{equation}
\label{3.10b}
p^*(t)=\frac{v^0}{2}(t-T)\cos \omega t -\frac{v^0}{2\omega }\sin \omega t.
\end{equation}
Now the integral in \eqref{3.10} can easily be computed giving 
\begin{equation}
\label{3.11}
\int_0^T (p^*\varphi ^*+\frac{1}{2}\varphi ^{*2}) \, dt=\int_0^{2\pi /\omega } \frac{(v^0)^2}{2\omega } (t-T)\cos \omega t \sin \omega t\, dt=-\frac{\pi (v^0)^2}{4\omega ^3}<0.
\end{equation}
Thus, $a^*=l/2$ is indeed the optimal solution of the problem. The chosen parameter enables the springs to have equal potential energy during the vibration of the oscillator. This turns out to maximize the potential action of the springs over one period of vibration among all admissible parameters and motions satisfying constraint \eqref{3.2}. Simultaneously the support force in O vanishes and a light and durable design is enabled.

\begin{figure}[htbp]
	\centering
		\includegraphics{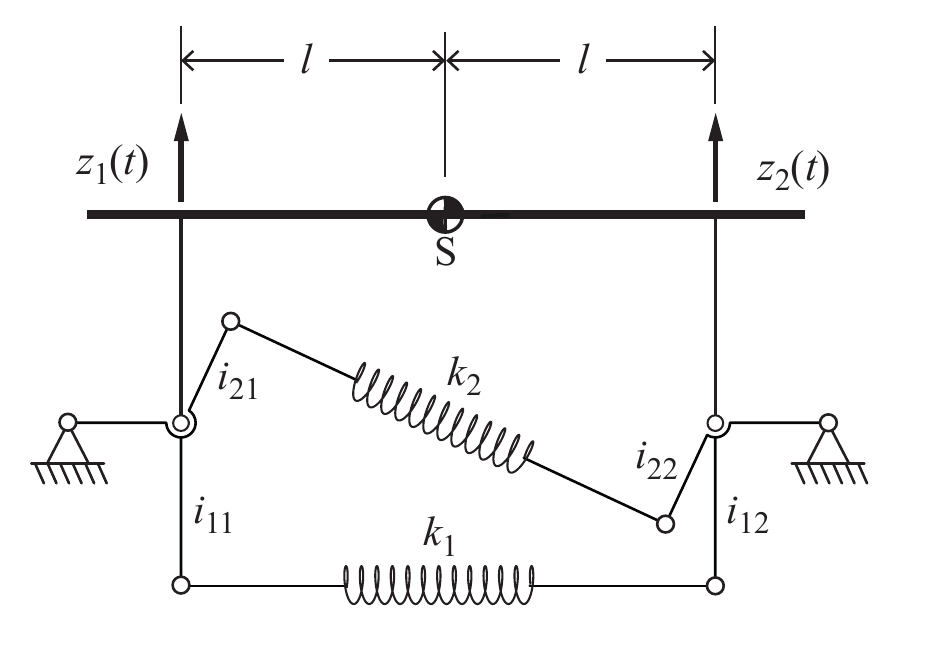}
	\caption{Model of a conservative interconnected suspension}
	\label{fig2}
\end{figure}

The next example deals with a model of interconnected suspension having two degrees of freedom as shown in Fig.~\ref{fig2} \citep{Neerpasch1997,Schulz2011}. In contrast to the previous example where we control the force application point through parameter $a$, we now control the influence of the springs on the vibration modes through four motion ratios $i_{11}$, $i_{12}$, $i_{21}$, and $i_{22}$ of the rockers. In this model the kinetic and potential energies of the oscillator are given by
\begin{align}
K(\dot{z}_1,\dot{z}_2)&=\frac{1}{2}m\left( \frac{\dot{z}_1+\dot{z}_2}{2}\right)^2+\frac{1}{2}\frac{mr^2}{l^2}\left( \frac{\dot{z}_1-\dot{z}_2}{2}\right)^2, \notag \\
U(z_1,z_2)&=\frac{1}{2}k_1(z_1i_{11}+z_2i_{12})^2+\frac{1}{2}k_2(z_1i_{21}+z_2i_{22})^2, \label{3.12}
\end{align}
with $m$ and $mr^2$ denoting the mass and moment of inertia (about S) of the beam, respectively, and $k_1$ and $k_2$ being the spring stiffnesses. We let $\mathbf{a}=(i_{11},i_{12},i_{21},i_{22})$ denote the parameter-vector controlling the action of the springs. Similar to the first example, where the applied restoring moment per unit displacement was limited by the fixed model parameters $(k, l)$, we now impose the following kinematic  constraints on $\mathbf{a}$
\begin{equation} \label{3.14} 
i_{11}^2+i_{12}^2=1/2, \quad
i_{21}^2+i_{22}^2=1/2. 
\end{equation}
As the magnitude of motion ratios cannot exceed $1/\sqrt{2}$, these kinematic constraints limit the magnitude of spring forces per unit displacements.

The first observation that can be made in connection with the spring energies given by \eqref{3.12} is that for the given displacements $z_1$ and $z_2$ the energy of the first (second) spring becomes maximal under the constraint \eqref{3.14} if $i_{11}/i_{12}=z_1/z_2$ ($i_{21}/i_{22}=z_1/z_2$). The simplest way to show this is to get rid of constraint \eqref{3.14} by introducing the parameter $\vartheta $ such that
\begin{displaymath}
i_{11}=\frac{1}{\sqrt{2}}\cos \vartheta ,\quad i_{12}=\frac{1}{\sqrt{2}}\sin \vartheta .
\end{displaymath}
Then the energy of the first spring becomes a function of $\vartheta $
\begin{equation}
\label{3.15}
U_1=\frac{1}{2}k_1(z_1i_{11}+z_2i_{12})^2=\frac{1}{4}k_1(z_1\cos \vartheta +z_2 \sin \vartheta )^2.
\end{equation}
This periodic function, defined for $\vartheta \in (0,2\pi )$, has a minimum achieved at $i_{12}/i_{11}=-z_1/z_2$ and a maximum achieved at $i_{11}/i_{12}=z_1/z_2$. The proof for the second spring is quite similar. Since there are two modes of vibration described by $\frac{1}{2}(z_1+z_2)$  (translation) and $\frac{1}{2}(z_1-z_2)$ (rotation), the above observation suggests that the energy of the first (second) spring becomes maximal when it acts against the translational (rotational) vibration.
 
Based on this observation we introduce the canonical coordinates 
\begin{equation}
\label{3.16}
\xi _1=\frac{1}{2}(z_1+z_2), \quad \xi _2=\frac{1}{2}(z_1-z_2).
\end{equation}
In these coordinates the kinetic and potential energies of the oscillator become
\begin{align}
\label{3.17}
K(\dot{\xi }_1,\dot{\xi }_2)&=\frac{1}{2}m_1\dot{\xi }_1^2+\frac{1}{2}m_2\dot{\xi }_2^2,\\
U(\xi _1,\xi _2)&=\frac{1}{2}k_1[\xi _1(i_{11}+i_{12})+\xi _2(i_{11}-i_{12})]^2 \notag
\\
&+\frac{1}{2}k_2[\xi _1(i_{21}+i_{22})+\xi _2(i_{21}-i_{22})]^2. \label{3.18}
\end{align}
With $\boldsymbol{\xi }(t)=(\xi _1(t),\xi _2(t))^T$, the equations of small vibrations, obtained by the energy method \citep{le2014energy}, read
\begin{equation}
\mathbf{M}\ddot{\boldsymbol{\xi }}(t)+\mathbf{K}(\mathbf{a})\boldsymbol{\xi }(t)=\mathbf{0}, \label{3.20} 
\end{equation}
where the mass matrix $\mathbf{M}=\text{diag}(m_1,m_2)$ is diagonal and does not depend on the control parameters, while the stiffness matrix $\mathbf{K}$ is given by
\begin{equation}
\label{3.20b}
\mathbf{K}=\begin{pmatrix}
 k_1(i_{11}+i_{12})^2+k_2(i_{21}+i_{22})^2     & k_1(i_{11}^{2}-i_{12}^{2})+k_2(i_{21}^{2}-i_{22}^{2})   \\
k_1(i_{11}^{2}-i_{12}^{2})+k_2(i_{21}^{2}-i_{22}^{2})     &  k_1(i_{11}-i_{12})^2+k_2(i_{21}-i_{22})^2
\end{pmatrix}.
\end{equation}
Equation \eqref{3.20} is subjected to the initial conditions
\begin{equation}
\label{3.20a}
\boldsymbol{\xi }(0)=0, \quad \dot{\boldsymbol{\xi }}(0)=\mathbf{v}^0.
\end{equation}
Our aim is to find the optimal motion ratios maximizing the potential action of the springs in one period of vibration
\begin{equation}
\int_0^T \{ \frac{1}{2}k_1[\xi _1(i_{11}+i_{12})+\xi _2(i_{11}-i_{12})]^2 
+\frac{1}{2}k_2[\xi _1(i_{21}+i_{22})+\xi _2(i_{21}-i_{22})]^2 \} dt \label{3.21}
\end{equation}
among all admissible $i_{11}$, $i_{12}$, $i_{21}$, $i_{22}$, $\xi _1$, and $\xi _2$ satisfying the constraints \eqref{3.14}, \eqref{3.20}, \eqref{3.20a}. 

To solve this optimization problem we introduce the control theory Lagrangian
\begin{multline}
L(\boldsymbol{\xi },\ddot{\boldsymbol{\xi }},\mathbf{p},\mathbf{a})=\mathbf{p}\cdot [\mathbf{M}\ddot{\boldsymbol{\xi }}+\mathbf{K}(\mathbf{a})\boldsymbol{\xi }] +\frac{1}{2}k_1[\xi _1(i_{11}+i_{12})+\xi _2(i_{11}-i_{12})]^2 
\\
+\frac{1}{2}k_2[\xi _1(i_{21}+i_{22})+\xi _2(i_{21}-i_{22})]^2. \label{3.22}
\end{multline}
The optimal solution $i_{11}^*$, $i_{12}^*$, $i_{21}^*$, $i_{22}^*$, $\xi _1^*(t)$, $\xi _2^*(t)$, $p _1^*(t)$, $p_2^*(t)$ must be found from equations \eqref{3.14},  \eqref{3.20}, \eqref{3.20a}, in which the unknown functions and control parameters must be replaced by those labelled with star. Furthermore the adjoint equation
\begin{equation}
\mathbf{M}\ddot{\mathbf{p}}^*(t)+\mathbf{K}(\mathbf{a}^*)\mathbf{p}^*(t)=-\mathbf{K}(\mathbf{a}^*)\boldsymbol{\xi }^*, \label{3.24} 
\end{equation}
for $\mathbf{p}^*(t)$ subjected to the end conditions
\begin{equation}
\label{3.25}
\mathbf{p}^*(T)=\dot{\mathbf{p}}^*(T)=0,
\end{equation}
must be fulfilled. Removing the term $\mathbf{p}\cdot \mathbf{M}\ddot{\boldsymbol{\xi }}$ in $L$ as independent of $\mathbf{a}$, we may present $F=\int Ldt$ as $k_1F_1+k_2F_2$, where
\begin{multline}
\label{3.25a}
F_1=\int_0^T \{ p_1^*[(i_{11}+i_{12})^2\xi _1^*+(i_{11}^2-i_{12}^2)\xi _2^*]+p_2^*[(i_{11}^2-i_{12}^2)\xi _1^*+(i_{11}-i_{12})^2\xi _2^*] 
\\
+ \frac{1}{2}[\xi _1^*(i_{11}+i_{12})+\xi _2^*(i_{11}-i_{12})]^2 \} dt
\end{multline}
is independent of $(i_{21},i_{22})$, and
\begin{multline}
\label{3.25b}
F_2=\int_0^T \{ p_1^*[(i_{21}+i_{22})^2\xi _1^*+(i_{21}^2-i_{22}^2)\xi _2^*]+p_2^*[(i_{21}^2-i_{22}^2)\xi _1^*+(i_{21}-i_{22})^2\xi _2^*] 
\\
+ \frac{1}{2}[\xi _1^*(i_{21}+i_{22})+\xi _2^*(i_{21}-i_{22})]^2 \} dt
\end{multline}
is independent of $(i_{11},i_{12})$. Thus, the maximization of $F$ with respect to $\mathbf{a}$ reduces to the two independent maximizations of $F_1$ and $F_2$ among the pairs $(i_{11},i_{12})$ and $(i_{21},i_{22})$ satisfying constraints \eqref{3.14}, respectively. 

We want to show that the optimal control parameters of this oscillator are 
\begin{equation}
\label{3.30}
i^*_{11}=i^*_{12}=\frac{1}{2},\quad i^*_{21}=-i^*_{22}=\frac{1}{2}.
\end{equation}
Indeed, for these chosen parameters the constraints \eqref{3.14} are satisfied identically. Next, the vector equation \eqref{3.20} becomes two uncoupled scalar equations
\begin{equation}
m_j\ddot{\xi}^*_j(t)+k_j\xi ^*_j(t)=0, \quad
j=1,2. \label{3.31}
\end{equation}
Each of these equations possesses the solution in the form
\begin{equation}
\label{3.33}
\xi ^*_j(t)=\frac{v^0_j}{\omega _j} \sin \omega _jt, \quad j=1,2,
\end{equation}
where $\omega _j$ are the eigenfrequencies of these normal modes of vibration
\begin{equation}
\label{3.34}
\omega _j=\sqrt{\frac{k_j}{m_j}}, \quad j=1,2.
\end{equation}
The vector equation \eqref{3.24} becomes also two uncoupled scalar equations
\begin{equation}
m_1\ddot{p}^*_j(t)+k_jp^*_j(t)=-k_j\xi ^*_j(t), \quad
j=1,2. \label{3.35}
\end{equation}
Substituting $\xi ^*_j(t)$ from \eqref{3.33} into \eqref{3.35} we find the solutions satisfying the end conditions \eqref{3.25} in the form
\begin{equation}
\label{3.36}
p^*_j(t)=\frac{v^0_j}{2}(t-T_j)\cos \omega _jt-\frac{v^0_j}{2\omega _j}\sin \omega _jt , 
\quad j=1,2.
\end{equation}
For $\omega _1\ne \omega _2$, both $\xi ^*_1(t)$ and $\xi ^*_2(t)$ will be periodic functions with the same period in three cases
\begin{itemize}
  \item[(i)] $v^0_1\ne 0$, $v^0_2=0$, $\xi ^*_1(t)\in $\eqref{3.33}, $T=T_1=2\pi /\omega _1$, $p^*_1(t)\in $\eqref{3.36}, $\xi ^*_2(t)=0$, $p^*_2(t)=0$,
  \item[(ii)] $v^0_2\ne 0$, $v^0_1=0$, $\xi ^*_2(t)\in $\eqref{3.33}, $T=T_2=2\pi /\omega _2$, $p^*_2(t)\in $\eqref{3.36}, $\xi ^*_1(t)=0$, $p^*_1(t)=0$,
  \item[(iii)] the frequency ratio $\omega _1/\omega _2$ is a rational number, then both $v^0_1$ and $v^0_2$ may be non-zero, and $T$ must be chosen as some common multiple of two periods $T_1=2\pi /\omega _1$ and $T_2=2\pi /\omega _2$.
\end{itemize}

Now, we turn to the necessary and sufficient conditions for the maximum of $F_1$ and $F_2$. Due to the constraint \eqref{3.14}$_1$, $(i_{11},i_{12})$ must lye on the circle of radius $1/\sqrt{2}$ in the $(i_{11},i_{12})$-plane. The unit tangential vector to this circle at point $(i_{11},i_{12})=(1/2,1/2)$ is $(-1,1)$. Thus, the derivative of $F_1$ in the tangential direction is
\begin{equation}
\label{3.36a}
\frac{dF_1}{ds}=\frac{\partial F_1}{\partial i_{12}}-\frac{\partial F_1}{\partial i_{11}},
\end{equation}
where $s$ is the arc-length along this circle. For the maximum of $F_1$ on the circle it is necessary that $dF_1/ds=0$. For $F_1$ from \eqref{3.25a} the derivative $dF_1/ds$ evaluated at point $(i_{11},i_{12})=(1/2,1/2)$ equals
\begin{equation}
\label{3.37}
\left. \frac{dF_1}{ds}\right|_{(1/2,1/2)}=-2\int_0^T [p^*_1(t)\xi ^*_2(t)+p^*_2(t)\xi ^*_1(t)+\xi ^*_1(t)\xi ^*_2(t)]\, dt.
\end{equation}
Similarly, at point $(i_{21},i_{22})=(1/2,-1/2)$ the unit tangential vector to the second circle is $(1,1)$, so the derivative of $dF_2/ds$, with $s$ being the same arc-length along the second circle, evaluated at that point equals
\begin{equation}
\label{3.37a}
\left. \frac{dF_2}{ds}\right|_{(1/2,-1/2)}=2\int_0^T [p^*_1(t)\xi ^*_2(t)+p^*_2(t)\xi ^*_1(t)+\xi ^*_1(t)\xi ^*_2(t)]\, dt.
\end{equation}
Thus, both necessary conditions $dF_1/ds=0$ and $dF_2/ds=0$ are equivalent. It is easy to see that they are fulfilled for each of the cases (i) or (ii). In case (iii) we may assume without restricting generality that $\omega _1=n_1\omega _0$, $\omega _2=n_2\omega _0$, and $T=2\pi /\omega _0$, where $n_1$ and $n_2$ are {\it different} natural numbers. Changing to the dimensionless time $\tau =\omega _0t$, we rewrite the last condition in the form
\begin{equation}
\label{3.38}
\int_0^{2\pi }[p^*_1(\tau )\xi ^*_2(\tau )+p^*_2(\tau )\xi ^*_1(\tau )+\xi ^*_1(\tau )\xi ^*_2(\tau )]\, d\tau =0.
\end{equation} 
In equation \eqref{3.38} the integral over $\tau $ of the products of $\sin n_1\tau $ and $\sin n_2\tau $ vanishes. Thus, it remains to show that
\begin{equation}
\label{3.39}
\int_0^{2\pi }[\frac{v^0_1}{2}(\tau -2\pi )\cos n_1\tau \frac{v^0_2}{n_2} \sin n_2\tau +\frac{v^0_2}{2}(\tau -2\pi )\cos n_2\tau \frac{v^0_1}{n_1} \sin n_1\tau ]\, d\tau =0.
\end{equation}
Using the identity $\cos n_1\tau d\tau =d(\sin n_1\tau )/n_1$ we integrate the first term by parts. Taking into account the initial and end conditions, we reduce both term to the product of sinus functions, whose integral vanishes. Thus, condition \eqref{3.37} is proved.

To verify that \eqref{3.30} maximizes the potential action over the period $T$, we need to consider the second derivatives of $F_1$ and $F_2$ along the circles evaluated at point $(i_{11},i_{12})=(1/2,1/2)$ and $(i_{21},i_{22})=(1/2,-1/2)$, respectively. As the tangential vector to the first circle at point $(i_{11},i_{12})=(1/2,1/2)$ is $(-1,1)$, we have
\begin{equation}
\label{3.40}
\left. \frac{d^2F_1}{ds^2}\right|_{(1/2,1/2)}=\frac{\partial ^2F_1}{\partial i_{11}^2}+\frac{\partial ^2F_1}{\partial i_{12}^2}-2\frac{\partial ^2F_1}{\partial i_{11}\partial i_{12}}.
\end{equation}
Similarly, the tangential vector to the second circle at point $(i_{21},i_{22})=(1/2,-1/2)$ is $(1,1)$, so the second derivative $d^2F_2/ds^2$ evaluated at that point is
\begin{equation}
\label{3.41}
\left. \frac{d^2F_2}{ds^2}\right|_{(1/2,-1/2)}=\frac{\partial ^2F_2}{\partial i_{21}^2}+\frac{\partial ^2F_2}{\partial i_{22}^2}+2\frac{\partial ^2F_2}{\partial i_{21}\partial i_{22}}.
\end{equation}
Computing these second derivatives with $F_1$ and $F_2$ from \eqref{3.25a} and \eqref{3.25b} and taking into account \eqref{3.37} and \eqref{3.37a}, we obtain
\begin{equation}
\label{3.42}
\frac{d^2F}{ds^2}=k_2 \left. \frac{d^2F_2}{ds^2}\right|_{(1/2,-1/2)}=4k_2\int_0^{T_1} (2p_1^*\xi _1^*+\xi _1^{*2}) \, dt=-2k_2\frac{\pi (v^0_1)^2}{\omega _1^3}<0
\end{equation}
in case (i), and
\begin{equation}
\label{3.43}
\frac{d^2F}{ds^2}=k_1 \left. \frac{d^2F_1}{ds^2}\right|_{(1/2,1/2)}=4k_1\int_0^{T_2} (2p_2^*\xi _2^*+\xi _2^{*2}) \, dt=-2k_1\frac{\pi (v^0_2)^2}{\omega _2^3}<0
\end{equation}
in case (ii).  In case (iii) we assume again that $\omega _1=n_1\omega _0$, $\omega _2=n_2\omega _0$ so that $T=2\pi /\omega _0$. Changing to the dimensionless time $\tau = \omega _0t$, we find the second derivative of $F$ to be
\begin{multline}
\label{3.44}
\frac{d^2F}{ds^2}=k_1 \left. \frac{d^2F_1}{ds^2}\right|_{(1/2,1/2)}+k_2 \left. \frac{d^2F_2}{ds^2}\right|_{(1/2,-1/2)}
\\
=\int_0^{2\pi } (8k_1\frac{v^0_2}{2}(\tau -2\pi )\cos n_2\tau \frac{v^0_2}{n_2} \sin n_2\tau  +8k_2\frac{v^0_1}{2}(\tau -2\pi )\cos n_1\tau \frac{v^0_1}{n_1} \sin n_1\tau)  \, d\tau 
\\
=-2k_1 \frac{(v^0_2)^2}{n_2}-2k_2 \frac{(v^0_1)^2}{n_1}<0.
\end{multline}
As the second derivatives are negative, function $F$ achieves its maximum at \eqref{3.30} in all three cases. These optimally chosen parameters enable the springs to act against the pure heave and roll modes of vibration of the beam separately, avoiding the energy transfer between different modes of vibrations.

\section{Application to dissipative oscillators}

\begin{figure}[htbp]
	\centering
		\includegraphics{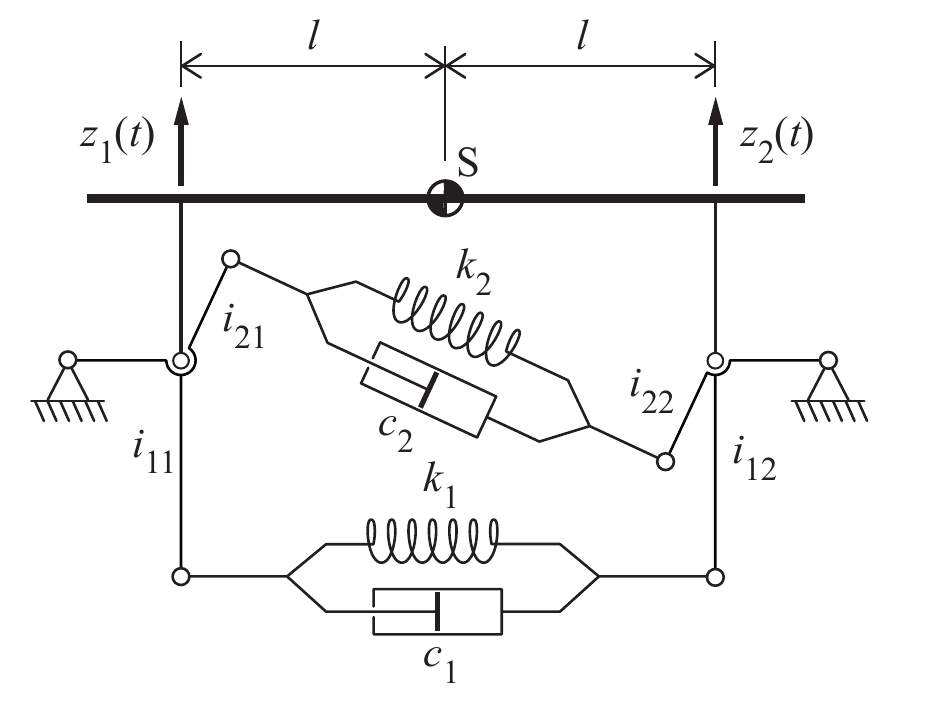}
	\caption{Model of dissipative interconnected suspension}
	\label{fig4}
\end{figure}

Consider now the model of interconnected suspension with two degrees of freedom as in the previous Section, but now with two dampers being included as shown in Fig.~\ref{fig4}. The kinetic and potential energies of the oscillator are given as before by \eqref{3.12}, but now, due to the dampers, we have also the dissipation function
\begin{equation}
\label{4.1}
D(\dot{z}_1,\dot{z}_2)=\frac{1}{2}c_1(\dot{z}_1i_{11}+\dot{z}_2i_{12})^2+\frac{1}{2}c_2(\dot{z}_1i_{21}+\dot{z}_2i_{22})^2,
\end{equation}
with $c_1$ and $c_2$ being the damping constants of the dampers. The parameters controlling the damped vibration of the beam are as before four motion ratios $i_{11}$, $i_{12}$, $i_{21}$, and $i_{22}$ of the rockers. We let $\mathbf{a}=(i_{11},i_{12},i_{21},i_{22})$  denote these parameters and assume that they satisfy the constraints \eqref{3.14}. Similar observation can be made in connection with the dissipation of dampers given by the above equation \eqref{4.1} is that for the given velocities $\dot{z}_1$ and $\dot{z}_2$ the dissipation potential of the first (second) damper becomes maximal under the constraint \eqref{3.14} if $i_{11}/i_{12}=z_1/z_2$ ($i_{21}/i_{22}=z_1/z_2$). Thus, we may expect that the dissipation of the first (second) damper becomes maximal when it acts against the translational (rotational) vibration.

Based on the above observation we use now the canonical coordinates \eqref{3.16} to write the kinetic and potential energies of the oscillator in form \eqref{3.17} and \eqref{3.18}, and the dissipation function in the form
\begin{align}
D(\dot{\xi }_1,\dot{\xi }_2)&=\frac{1}{2}c_1[\dot{\xi }_1(i_{11}+i_{12})+\dot{\xi }_2(i_{11}-i_{12})]^2 \notag
\\
&+\frac{1}{2}c_2[\dot{\xi }_1(i_{21}+i_{22})+\dot{\xi }_2(i_{21}-i_{22})]^2. \label{4.2}
\end{align}
Consequently, the equations of small vibrations of this oscillator, obtained by the energy method \citep{le2014energy}, read
\begin{equation}
\mathbf{M}\ddot{\boldsymbol{\xi }}(t)+\mathbf{C}(\mathbf{a})\dot{\boldsymbol{\xi }}(t)+\mathbf{K}(\mathbf{a})\boldsymbol{\xi }(t)=\mathbf{0}, \label{4.4} 
\end{equation}
where $\mathbf{M}$ and $\mathbf{K}$ remain the same as in the previous example, while the damping matrix $\mathbf{C}$ is given by
\begin{equation}
\label{4.5}
\mathbf{C}=\begin{pmatrix}
 c_1(i_{11}+i_{12})^2+c_2(i_{21}+i_{22})^2     & c_1(i_{11}^{2}-i_{12}^{2})+c_2(i_{21}^{2}-i_{22}^{2})   \\
c_1(i_{11}^{2}-i_{12}^{2})+c_2(i_{21}^{2}-i_{22}^{2})     &  c_1(i_{11}-i_{12})^2+c_2(i_{21}-i_{22})^2
\end{pmatrix}.
\end{equation}
Equation \eqref{4.4} is subjected to the initial conditions \eqref{3.20a}. Our aim is to find the optimal motion ratios maximizing the potential action of the springs
\begin{equation}
\int_0^T \{ \frac{1}{2}k_1[\xi _1(i_{11}+i_{12})+\xi _2(i_{11}-i_{12})]^2 
+\frac{1}{2}k_2[\xi _1(i_{21}+i_{22})+\xi _2(i_{21}-i_{22})]^2 \} dt \label{4.7}
\end{equation}
among all admissible parameters $i_{11}$, $i_{12}$, $i_{21}$, $i_{22}$, $\xi _1$, and $\xi _2$ satisfying the constraints \eqref{3.14}, \eqref{4.4}, \eqref{3.20a}.

To solve this optimization problem we introduce the control theory Lagrangian
\begin{multline}
L(\boldsymbol{\xi },\dot{\boldsymbol{\xi }},\ddot{\boldsymbol{\xi }},\mathbf{p},\mathbf{a})=\mathbf{p}\cdot [\mathbf{M}\ddot{\boldsymbol{\xi }}+\mathbf{C}(\mathbf{a})\dot{\boldsymbol{\xi }}+\mathbf{K}(\mathbf{a})\boldsymbol{\xi }] +\frac{1}{2}k_1[\xi _1(i_{11}+i_{12})+\xi _2(i_{11}-i_{12})]^2 
\\
+\frac{1}{2}k_2[\xi _1(i_{21}+i_{22})+\xi _2(i_{21}-i_{22})]^2.  \label{4.8}
\end{multline}
The optimal solution $i_{11}^*$, $i_{12}^*$, $i_{21}^*$, $i_{22}^*$, $\xi _1^*(t)$, $\xi _2^*(t)$, $p _1^*(t)$, $p_2^*(t)$ must be found from equations \eqref{3.14},  \eqref{4.4}, \eqref{3.20a}, in which the unknown functions and control parameters must be replaced by those labelled with star. Furthermore the adjoint equation
\begin{equation}
\mathbf{M}\ddot{\mathbf{p}}^*(t)-\mathbf{C}(\mathbf{a}^*)\dot{\mathbf{p}}^*(t)+\mathbf{K}(\mathbf{a}^*)\mathbf{p}^*(t)=-\mathbf{K}(\mathbf{a}^*)\boldsymbol{\xi }^*(t), \label{4.9} 
\end{equation} 
subjected to the end conditions
\begin{equation}
\label{4.9a}
\dot{\mathbf{p}}^*(T)=\dot{\mathbf{p}}^*(T)=0,
\end{equation}
must be fulfilled for $\mathbf{p}^*(t)=(p^*_1(t),p^*_2(t))^T$.  The maximization of $\int Ldt$ with respect to $\mathbf{a}$ reduces to the two independent maximizations of
\begin{multline}
\label{4.10}
F_1=\int_0^T \{ p_1^*[c_1(i_{11}+i_{12})^2\dot{\xi }_1^*+c_1(i_{11}^2-i_{12}^2)\dot{\xi }_2^*]+p_2^*[c_1(i_{11}^2-i_{12}^2)\dot{\xi }_1^* 
\\
+c_1(i_{11}-i_{12})^2\dot{\xi }_2^*]+p_1^*[k_1(i_{11}+i_{12})^2\xi _1^*+k_1(i_{11}^2-i_{12}^2)\xi _2^*]+p_2^*[k_1(i_{11}^2-i_{12}^2)\xi _1^*
\\
+k_1(i_{11}-i_{12})^2\xi _2^*]+ \frac{1}{2}k_1[\xi _1^*(i_{11}+i_{12})+\xi _2^*(i_{11}-i_{12})]^2 \} dt
\end{multline}
and
\begin{multline}
\label{4.11}
F_2=\int_0^T \{ p_1^*[c_2(i_{21}+i_{22})^2\dot{\xi }_1^*+c_2(i_{21}^2-i_{22}^2)\dot{\xi }_2^*]+p_2^*[c_2(i_{21}^2-i_{22}^2)\dot{\xi }_1^*
\\
+c_2(i_{21}-i_{22})^2\dot{\xi }_2^*] +p_1^*[k_2(i_{21}+i_{22})^2\xi _1^*+k_2(i_{21}^2-i_{22}^2)\xi _2^*]+p_2^*[k_2(i_{21}^2-i_{22}^2)\xi _1^*
\\
+k_2(i_{21}-i_{22})^2\xi _2^*]+ \frac{1}{2}k_2[\xi _1^*(i_{21}+i_{22})+\xi _2^*(i_{21}-i_{22})]^2 \} dt
\end{multline}
among the pairs $(i_{11},i_{12})$ and $(i_{21},i_{22})$ satisfying constraints \eqref{3.14}, respectively. In \eqref{4.10} and \eqref{4.11} the term $\mathbf{p}\cdot \mathbf{M}\ddot{\boldsymbol{\xi }}$ is neglected as independent of $\mathbf{a}$. 

We want to show that the optimal control parameters of this dissipative oscillator are 
\begin{equation}
\label{4.14}
i^*_{11}=i^*_{12}=\frac{1}{2},\quad i^*_{21}=-i^*_{22}=\frac{1}{2}.
\end{equation}
Indeed, for these chosen parameters the constraints \eqref{3.14} are satisfied identically. Next, the vector equation \eqref{4.4} becomes two uncoupled scalar equations
\begin{equation}
m_j\ddot{\xi}^*_j+c_j\dot{\xi}^*_j+k_j\xi ^*_j=0, \quad
j=1,2. \label{4.15}
\end{equation}
Each of \eqref{4.15} possesses non-trivial solution fulfilling one of \eqref{3.20a} in the form
\begin{equation}
\label{4.17}
\xi ^*_j(t)=\frac{v^0_j}{\nu _j}e^{-h_jt} \sin \nu _jt, \quad j=1,2,
\end{equation}
where 
\begin{equation}
\label{4.18}
h_j=\frac{c_j}{2m_j} , \quad \nu _j= \frac{\sqrt{4m_jk_j-c_j^2}}{2m_j}, \quad j=1,2,
\end{equation}
provided $c_j<2\sqrt{m_jk_j}$ (underdamped vibration modes). The vector equation \eqref{4.9} becomes also two uncoupled scalar equations
\begin{equation}
m_j\ddot{p}^*_j-c_j\dot{p}^*_j+k_jp^*_j=-k_1\xi ^*_j, \quad
j=1,2. \label{4.19}
\end{equation}
Substituting $\xi ^*_1(t)$ and $\xi ^*_2(t)$ from \eqref{4.17} into \eqref{4.19} we find the solutions satisfying the end conditions \eqref{4.9a} in the form
\begin{equation}
\label{4.20}
p^*_j(t)=-\frac{v^0_j}{4}[e^{-h_jt}(\frac{1}{h_j}\cos \nu _jt+\frac{1}{\nu _j}\sin \nu _jt)-e^{h_j(t-4\pi /\nu _j)}(\frac{1}{h_j}\cos \nu _jt -\frac{1}{\nu _j}\sin \nu _jt)].
\end{equation}
Since both $\xi ^*_1(t)$ and $\xi ^*_2(t)$ must simultaneously vanish at $t=0$ and $t=T$ to have the conditionally periodic vibrations, we analyze in the general case $\nu _1\ne \nu _2$ the following two solutions
\begin{itemize}
  \item[(i)] $\xi ^*_1(t)\in $\eqref{4.17}, $T=T_1=2\pi /\nu _1$, $p^*_1(t)\in $\eqref{4.20}, $\xi ^*_2(t)=p^*_2(t)=0$,
  \item[(ii)] $\xi ^*_2(t)\in $\eqref{4.17}, $T=T_2=2\pi /\nu _2$, $p^*_2(t)\in $\eqref{4.20}, $\xi ^*_1(t)=p^*_1(t)=0$.
\end{itemize}

We turn now to the necessary and sufficient conditions for the maximum of $F_1$ and $F_2$. As before, the constraint \eqref{3.14}$_1$ forces the point $(i_{11},i_{12})$ to lye on the circle of radius $1/\sqrt{2}$ in the $(i_{11},i_{12})$-plane. Therefore, for the maximum of $F_1$ at point $(i_{11},i_{12})=(1/2,1/2)$ it is necessary that
\begin{equation}
\label{4.20a}
\left. \frac{dF_1}{ds}\right|_{(1/2,1/2)}=-2\int_0^T [c_1(p^*_1\dot{\xi }^*_2+p^*_2\dot{\xi }^*_1)+k_1(p^*_1\xi ^*_2+p^*_2\xi ^*_1+\xi ^*_1\xi ^*_2)]\, dt=0.
\end{equation}
Similarly, the maximum of $F_2$ is achieved at point $(i_{21},i_{22})=(1/2,-1/2)$ if
\begin{equation}
\label{4.21}
\left. \frac{dF_2}{ds}\right|_{(1/2,-1/2)}=2\int_0^T [c_2(p^*_1\dot{\xi }^*_2+p^*_2\dot{\xi }^*_1)+k_2(p^*_1\xi ^*_2+p^*_2\xi ^*_1+\xi ^*_1\xi ^*_2)]\, dt=0.
\end{equation}
Since $k_1\ne k_2$ and $c_1\ne c_2$, conditions \eqref{4.20a} and \eqref{4.21} are satisfied if
\begin{equation}
\label{4.22}
\int_0^T [p^*_1(t)\xi ^*_2(t)+p^*_2(t)\xi ^*_1(t)+\xi ^*_1(t)\xi ^*_2(t)]\, dt=0,
\end{equation}
and
\begin{equation}
\label{4.23}
\int_0^T [p^*_1(t)\dot{\xi }^*_2(t)+p^*_2(t)\dot{\xi }^*_1(t)]\, dt=0.
\end{equation}
Thus, it is easy seen that both necessary conditions \eqref{4.22} and \eqref{4.23} are fulfilled for each of the cases (i) or (ii).  

Finally, we need to check the necessary conditions that the second derivatives of $F_1$ and $F_2$ in the tangential directions to the circles evaluated at $(i_{11},i_{12})=(1/2,1/2)$ and  $(i_{21},i_{22})=(1/2,-1/2)$ are negative. Computing these second derivatives with $F_1$ and $F_2$ from \eqref{4.10} and \eqref{4.11} and taking into account \eqref{4.22} and \eqref{4.23}, we obtain
\begin{equation}
\label{4.24}
\frac{d^2F}{ds^2}=\left. \frac{d^2F_2}{ds_2^2}\right|_{(1/2,-1/2)}=4\int_0^T [k_2(2p_1^*\xi _1^*+\xi _1^{*2})+2c_2p_1^*\dot{\xi }_1^*] \, dt
\end{equation}
in case (i), and
\begin{equation}
\label{4.25}
\frac{d^2F}{ds^2}=\left. \frac{d^2F_1}{ds^2}\right|_{(1/2,1/2)}=4\int_0^T [k_1(2p_2^*\xi _2^*+\xi _2^{*2})+2c_1p_2^*\dot{\xi }_2^*] \, dt
\end{equation}
in case (ii). The integrals standing on the right-hand sides give
\begin{equation}
\label{4.26}
\frac{d^2F}{ds^2}=-\frac{(v^0_1)^2}{2h_1^2}\left[ c_2+e^{-2h_1T_1}\frac{4\pi k_2h_1^2-c_2(4\pi h_1^3+4\pi h_1\nu _1^2+\nu _1^3)}{\nu _1^3}\right]
\end{equation}
in case (i), and
\begin{equation}
\label{4.27}
\frac{d^2F}{ds^2}=-\frac{(v^0_2)^2}{2h_2^2}\left[ c_1+e^{-2h_2T_2}\frac{4\pi k_1h_2^2-c_1(4\pi h_2^3+4\pi h_2\nu _2^2+\nu _2^3)}{\nu _2^3}\right]
\end{equation}
in case (ii). In the limit $c_1\to 0$, $c_2\to 0$, these expressions tend to \eqref{3.43} and \eqref{3.44} which are negative. If $c_1$ and $c_2$ are large, the terms with exponential factors are negligibly small, so the second derivatives are also negative. Therefore function $F$ achieves its maximum at \eqref{4.14} in both cases (i) and (ii). Thus, \eqref{4.14} represents the optimal control parameters of the oscillator. These optimally chosen parameters enable the springs to act against the pure heave and roll modes of vibration of the beam separately, avoiding the energy transfer between different modes of vibrations.

\section{Discussions and conclusions}

Let us first mention that the method used in this paper can be applied to optimize parameters controlling small vibrations of any oscillator having a finite number of degrees of freedom. If a concept of a dynamic system based on this optimization is built, we may expect that there is no energy transmission of springs or dampers between different modes of vibration. The system always stays uncoupled and simultaneously a tuning of the system with no compromises is enabled. Note in addition that this uncoupling of vibration modes improve also the controllability and accessibility of the oscillator, and consequently the further optimization of spring and damper characteristics to meet other goals can easily be provided. 

The other interesting observation is that the optimal motion ratios are independent of maximizing the potential action or dissipation function. If we take a more careful look in our optimization procedure, we may notice that the mass distribution was the only property which was not changed by this optimization approach. If we derive the eigenvectors of the mass matrix obtained from \eqref{3.12}
\begin{equation*}
\mathbf{M}=\frac{m}{4 \, l^2}\begin{pmatrix}
 l^2+r^2     	&				l^2-r^2   \\
l^2-r^2     		&				l^2+r^2
\end{pmatrix},
\end{equation*}
we get the coordinates which diagonalize (uncouple) our kinetic energy. It is obvious that the intuitively chosen canonical coordinates in \eqref{3.16} already contain the optimal solution. Physically these eigenvectors describe the central principal axis of inertia and the eigenvalues are equal to the principal moment of inertia of the rigid body.

For a dynamical system having $n$ degrees of freedom there exist always $n$ canonical coordinates. These canonical coordinates are only based on the mass allocation of the system and can be found through diagonalizing the kinetic energy (or mass matrix) before any dynamic concept with springs and dampers is build. These easily accessible parameters describe the perfect subdivision of an acting force in the related mode of vibration to control the system most effectively and maximize the potential energy or dissipation function. If these canonical coordinates are use to built a new dynamic concept the optimality is expected to be realized. Because such a design concept is just depending on the mass allocation of the system, we will call this design approach the Mass Allocation Character Approach (short: MAC-Approach).

In a forthcoming publication we will show, in the style of the rigid body example of this paper, that the optimal control based on the canonical coordinates does work for systems with more rigid bodies and in 3-D case. However in that case the principal axes of inertia turn out to be more fictive and not the simple axes.

\end{document}